\documentclass[twocolumn,fleqn]{article}
\usepackage{graphicx}
\title{An Analysis of 900 Optical Rotation Curves: \\
       The Universal Rotation Curve As A Power-Law \\
       And The Development Of A Theory-Independent\\
       Dark-Matter Modeller}
\author{D. F. Roscoe, School of Mathematics, \\
Sheffield University, Sheffield, S3 7RH, UK. \\
Email: D.Roscoe@shef.ac.uk \\
Tel: 0114-2223791, Fax: 0114-2223739}
\normalsize
\begin{document}
\maketitle
\abstract
One of the largest $H_\alpha$ rotation curve data bases of spiral galaxies
currently
available is that provided by Persic \& Salucci, (astro-ph/9502091) 
hereafter PS 1995, which has
been derived by them from unreduced rotation curve data of 965 southern sky
spirals obtained by Mathewson, Ford \& Buchhorn, hereafter MFB 1992.
Of the original sample of 965 galaxies, the observations on 900 were considered
by PS 1995 to be good enough for rotation curve studies, and the present
analysis concerns itself with these 900 rotation curves.

The analysis is performed within the context of the basic hypothesis that the
phenomenology of rotation curves in the optical disc (that is, away from the
dynamical effects of the bulge) can be systematically described in terms of a
general power-law $V = A\,R^\alpha$, valid for $R \, > \,R_{min}$, where
$R_{min}$ is an estimate of the transition radius between bulge-dominated
and disc-dominated dynamics.
The analysis begins by showing how this model provides an extremely good
description of the generic behaviour of rotation curves in the optical disc
and, furthermore, how it imposes very detailed correlations between the free
parameters, $A$ and $\alpha$, of the model.

These correlations are investigated, and shown to imply, via first and
second-order models, a third-order model according to which the rotation
velocity, $V$, at any radial displacement in the optical disc of any given
spiral galaxy is given by $V /V_0 = (R/R_0)^\alpha$, where $R_0 > R_{min}$, and
$V_0$ are given as approximate functions of the galaxy's absolute magnitude and
surface brightness whilst $\alpha$ is an unidentified function of other galaxy
parameters - of which the most significant ones will be the relative proportions
of the disc, bulge and halo mass-components.
It is this latter function which provides the opportunity for a dark-matter
modelling process which is independent of any particular dynamical theory.

Furthermore, it is shown that the conclusion of PS 1986, that optical-disc
dynamics contain no signature of the transition from disc-dominated dynamics
to halo-dominated dynamics, is extremely strongly supported by this analysis.
\section{Introduction}
\label{sec.1}
It has been know for several years - for example, Rubin, Burstein, Ford
\& Thonnard, hereafter RBFT - that optical rotation curve shapes are
strongly determined by luminosity in the sense that, within any given
morphological class and away from the innermost part of the rotation curve,
high luminosity galaxies have almost flat rotation curves, whilst low luminosity
galaxies have more {\lq rounded and rising'} rotation curves.

However, until recently all studies of the correlations existing between
rotation curve kinematics and global galaxy properties have necessarily
been confined to relatively small samples of rotation curves, and have
therefore been subject to the corresponding uncertainties.
This problem has been rectified by the publication of 900 good quality
rotation curves by PS 1995.
A first analysis of this data was given by Persic et al (1996), hereafter PSS,
in which they concluded
that there exists a {\lq universal rotation curve'} for spiral galaxies which
is determined primarily by the total luminosity of the galaxy concerned, and
the essence of this conclusion is confirmed here.
The present analysis of the same data differs crucially from that of PSS in that
an early decision was taken to analyse the data within the context of the
hypothesis that rotation curves in the optical disc (that is, away from the
dynamical effects of the bulge) could be reasonably described by a
generalized power law, $V_{rot} = A\, R^\alpha$, where $R\,>\,R_{min}$
and $R_{min}$ represents an estimate of the transition radius from
bulge-dominated to disc-dominated dynamics.
This decision, which was motivated by arguments which are independent of
the matters at hand, could easily have resulted in relative failure in the
sense of providing no particular insights into rotation curve structure.
In fact, the opposite happened and it was found that the hypothesis imposes
extremely strong correlations between rotation curve kinematics, total
luminosities and surface brightnesses, thereby giving a detailed confirmation
to the general thrust of the PS studies, and a strong confirmation to the
particular idea of the universal rotation curve.

In detail, the primary conclusion of this analysis is that the rotation curve
in the optical disc of any given spiral galaxy with absolute magnitude $M$ and
surface brightness $S$ behaves according to
\begin{eqnarray}
{ V \over V_0 } &=& \left( { R \over R_0 } \right)^\alpha, \label{1a} \\
\log V_0 &\approx& -0.584 - 0.133 \,M - 0.000243\, S, \nonumber \\
\log R_0 &\approx& -3.291 - 0.208 \,M - 0.00292\, S, \nonumber  \\
\alpha &\approx& g(d_m,\,b_m,\,h_m), \nonumber
\end{eqnarray}
where $g(d_m,\,b_m,\,h_m)$ is some undetermined function for which the most
significant parameters are probably the
disc-mass ($d_m$), bulge-mass ($b_m$) and halo-mass ($h_m$) respectively.
Since this model is shown to account for over 90\% of the variation in the
pivotal diagram (equivalent to a regression correlation of 0.95), it can be
considered as, at the very least, an extremely good approximation to the
statistical reality, as judged for a large number of spiral discs.

The best model that can be given for the undetermined exponent, $\alpha$, in
terms of the data provided by PS 1995 is that $\alpha=2.56+0.105\,M$, which
accounts for over 29\% of the variation in the $(\alpha,\,M)$ plot (equivalent
to a regression correlation of 0.55).
We see immediately from this model that high luminosity galaxies are
the ones which possess the flattest rotation curves, whilst the low luminosity
galaxies are the ones which possess the {\lq rounded and rising'} rotation
curves - thereby confirming the RBFT result for a very large sample.

A further, and very significant, consequence of the success of (\ref{1a}) is
its implied confirmation of the PS 1986 result that optical discs appear to
contain no signature of the transition from disc-dominated dynamics to
halo-dominated dynamics.
PS reached this conclusion after an analysis of 42 rotation curves showed the
near-constancy, over the whole of each optical rotation curve, of the parameter
${\hat V} \equiv (\Omega -K/2)R$, for angular velocity $\Omega$ and epicyclic
frequency $K$.
The connection between this result of PS and the present perspective is
discussed in Appendix \ref{app.1b}.

Further evidence supporting the apparent absence of any kinematical transition
region in the optical disc arises through the discovery of very strong
correlation between $R_{opt}$ and $R_{min}$, where $R_{min}$ is a
theory-independent estimate of the transition region between bulge-dominated
and disc-dominated dynamics (see Appendix \ref{app.4}).
The existence of such a correlation implies a causal connection between
$R_{min}$ and $R_{opt}$ which, again, argues against the existence of sharply
delineated transition regions between disc-dominated and halo-dominated
dynamics.

Finally, and of potentially greatest significance, it is to be noted that
halo-mass only enters the system through the undetermined function $\alpha$;
this fact provides the possibility of a dark-matter modelling process which is
independent of any particular dynamical theory.
Specifically, since, for any given galaxy, reasonable estimates for $d_m$ and
$b_m$ can be obtained, then it becomes possible to model $\alpha$ in terms of
these two quantities.
Analysis of any systematic differences arising between $\alpha$-values
calculated from the rotation curves, and the modelled $\alpha$-values,
will then, in principle, allow systematic estimates of any given galaxy's
dark-matter content to be given in terms of the galaxy's disc-mass, bulge-mass
and surface brightness.
\section{The Data}
\label{sec.2}
The data given by PS is obtained from the raw $H \alpha$ data of MFB by
deprojection, folding and cosmological redshift correction.
For any given galaxy, the data is presented in the form of estimated rotational
velocities plotted against angular displacement from the galaxy's centre;
estimated linear scales are not given and no data-smoothing is performed.

The analysis proposed here requires the linear scales of the galaxies in
the sample to be defined which, in turn, requires distance estimates of
the sample galaxies from our own locality.
This information is given in the original MFB paper in the form of three
variations of the Tully-Fisher, hereafter TF, distance estimate, say TF1, TF2
and TF3.
TF1 is derived by using the TF relation for Fornax obtained from magnitude
data; TF2 is the Malmquist-bias corrected form of TF1 giving
TF2=1.08$\times$TF1 for all galaxies; TF3 is obtained by using the TF relation
for Fornax obtained from rotation curve data.
The current analysis is based upon TF3.

Table 3 gives the morphological type-distribution in the PS 1995 data base,
and shows that the great majority of the selected galaxies are of types
3,4,5 and 6, with only two examples of types 0,1,2 and a tail of 31 examples
of types 7,8,9.
\begin{center}
\begin{tabular}{||r|r||}
\hline
\hline
\multicolumn{2}{||c||}{\bf Table 3} \\
\hline
\hline
Galaxy & Sample  \\
Type   & Size \\
\hline \hline
0,1,2 &   2 \\
3     & 306 \\
4,5   & 177 \\
6     & 384 \\
7,8,9 &  31 \\
\hline
\hline
\end{tabular}
\end{center}
For the purposes of the following analysis, no distinction is made between
type classes, and so the whole sample of 900 rotation curves is used.
However, the results of this analysis remain unchanged, except in the
numerical details, when it is repeated on any of the subsets consisting of
types $\left\{ 3 \right\}$, $\left\{ 4,5 \right\}$ or $\left\{ 6 \right\}$.

The details of the data-reduction used in these analyses are given in
Appendix \ref{app.4} and, briefly, they amount to a {\it prior-decided} means of
minimizing the effect of the bulge on the rotation-curve calculations.
\section{A Necessary Condition For \hfill \break
         The Power-Law Hypothesis}
\label{sec.3}
\begin{figure*}
\begin{center}
\includegraphics*[width=3.5in,keepaspectratio]{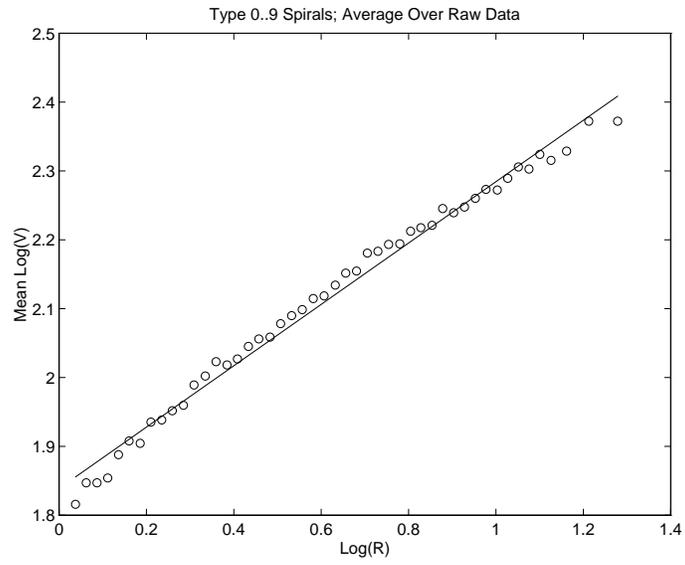}
\caption{Mean Of 900 Individual Rotation Curves}
\label{fig.0}
\end{center}
\end{figure*}

Suppose the power-law hypothesis of (\ref{1a}) to be true; in this case, and
excepting for the
inevitable high level of noise, a plot of any given rotation curve in the
$(\log R,\,\log V)$ plane would lie on a straight line.
Since the rotation curves are not identical then, for the 900 rotation
curves, we would obtain 900 distinct straight-line plots in the
$(\log R,\,\log V)$ plane.
In the following, we show that, in this case, the mean plot of the 900
separate plots must also be a straight line: For a given $\log R$, the
corresponding $\log V$ value for each of the 900 rotation curves would be
given by
\begin{displaymath}
\log V_k = a_k + b_k\, \log R,~~k=1..900.
\end{displaymath}
Summing over $k$, and forming the mean response, we find
\begin{displaymath}
{\rm Mean}\left( \log V \right) = A + B\, \log R,
\end{displaymath}
so that, as stated, the mean plot is also a straight line.
It follows that a necessary condition for optical rotation curve data to be
described by a power law is that the mean plot of all the rotation curves in the
$(\log R,\,\log V)$ plane must be a (statistically) straight line.
So, the strategy of the initial analysis is described as follows:
\begin{itemize}
\item
Reduce all the data in the combined sample of 900 rotation curves to a
uniform linear scale based on TF3;
\item
Superimpose all the data of the combined sample into a single data set;
\item
Divide the data set into bins subject to some convenient criteria.
For present purposes, the bin-width was chosen as $0.057kpc$ subject to
the constraint that each bin contained at least 200 data points.
In practice, the average per bin is about 800 data points.
\item
Form the average of the data in each bin and plot it.
\end{itemize}
The rationale is simply that, if the power law hypothesis is correct, then the
internal noise on the data for the 900 separate rotation curves should largely
cancel out over the averaging process.
The result is plotted in Figure \ref{fig.0} - and it is to be emphasized that
this plot does not represent a physical rotation curve, but is merely a
convenient graphic tool designed to illustrate a statistical point.
The only discernible sense of any possible systematic deviation from a purely
linear behaviour occurs in the innermost part of the plot;
apart from this, the figure provides solid support for the idea that a
general power law is, at the very least, a good working approximation to
$H_\alpha$ rotation curve data, and provides the necessary rationale for
the basic analysis of this paper.
\section{The Minimization Of Bulge-Dominated
         Dynamics In Rotation Curve Data}
Before beginning the analysis proper, we note that the apparent deviation from
linear behaviour on the innermost part of the rotation curve, noted in the
previous section, is easily understood in terms of the transition from
bulge-dominated to disc-dominated dynamics.
In this section, we describe the effect of implementing the data-reduction
process described in Appendix \ref{app.4} which is designed to minimize the
effects of bulge-dominated dynamics in the present considerations .
In effect, this process is a numerical process which removes the innermost
data points on any given rotation curve according to how statistically
{\lq unusual'} those points are in relation to the rest of the data, and
its effects are described below in Figures \ref{fig.1a} \& \ref{fig.1b} and
in Tables 1 \& 2.

Figure \ref{fig.1a} plots the measured $R_{min}$ against the measured
$R_{opt}$ {\it before} any data reduction is performed, whilst
Figure \ref{fig.1b} plots the {\it calculated} $R_{min}$ arising from the
data reduction process against the measured $R_{opt}$.
It is clear from these diagrams that the data reduction process reveals
a fairly strong correlation between $R_{min}$ and $R_{opt}$.
The change between the two diagrams is quantified in Tables 1 \& 2.
\begin{center}
\begin{tabular}{||l|c|c|r|c||}
\hline
\hline
\multicolumn{5}{||c||}{\bf Table 1}  \\
\hline
\hline
\multicolumn{5}{||c||}{$R_{opt} = b_0 + b_1 \,R_{min} $}   \\
\multicolumn{5}{||c||}{Before data reduction} \\
\hline
\hline
Predictor  & Coeff & Std Dev & t-ratio & p \\
\hline \hline
Const.    & 7.25 & 0.13  & 57 & 0.00 \\
$R_{min}$         & 1.62 &0.11  & 14 & 0.00 \\
\hline
\multicolumn{5}{||c||}{$R^2$ ~=~ 19.2\%} \\
\hline
\hline
\multicolumn{5}{||c||}{\bf Table 2}  \\
\hline
\hline
\multicolumn{5}{||c||}{After data reduction} \\
\hline
\hline
Predictor  & Coeff & Std Dev & t-ratio & p \\
\hline \hline
Const.    & 5.73 & 0.14  & 41 & 0.00 \\
$R_{min}$         & 1.64 &0.07  & 24 & 0.00 \\
\hline
\multicolumn{5}{||c||}{$R^2$ ~=~ 38.6\%} \\
\hline
\hline
\end{tabular}
\end{center}
\begin{figure*}
\begin{center}
\includegraphics*[width=3.5in,keepaspectratio]{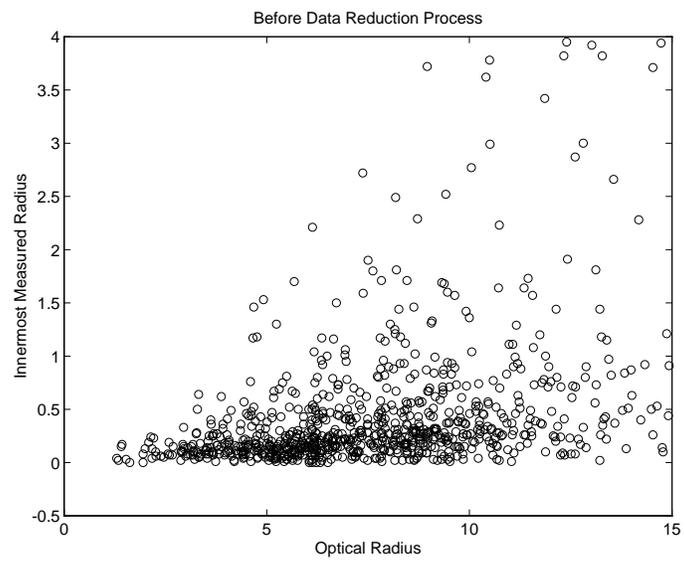}
\caption{Before Data Reduction}
\label{fig.1a}
\end{center}
\end{figure*}
\begin{figure*}
\begin{center}
\includegraphics*[width=3.5in,keepaspectratio]{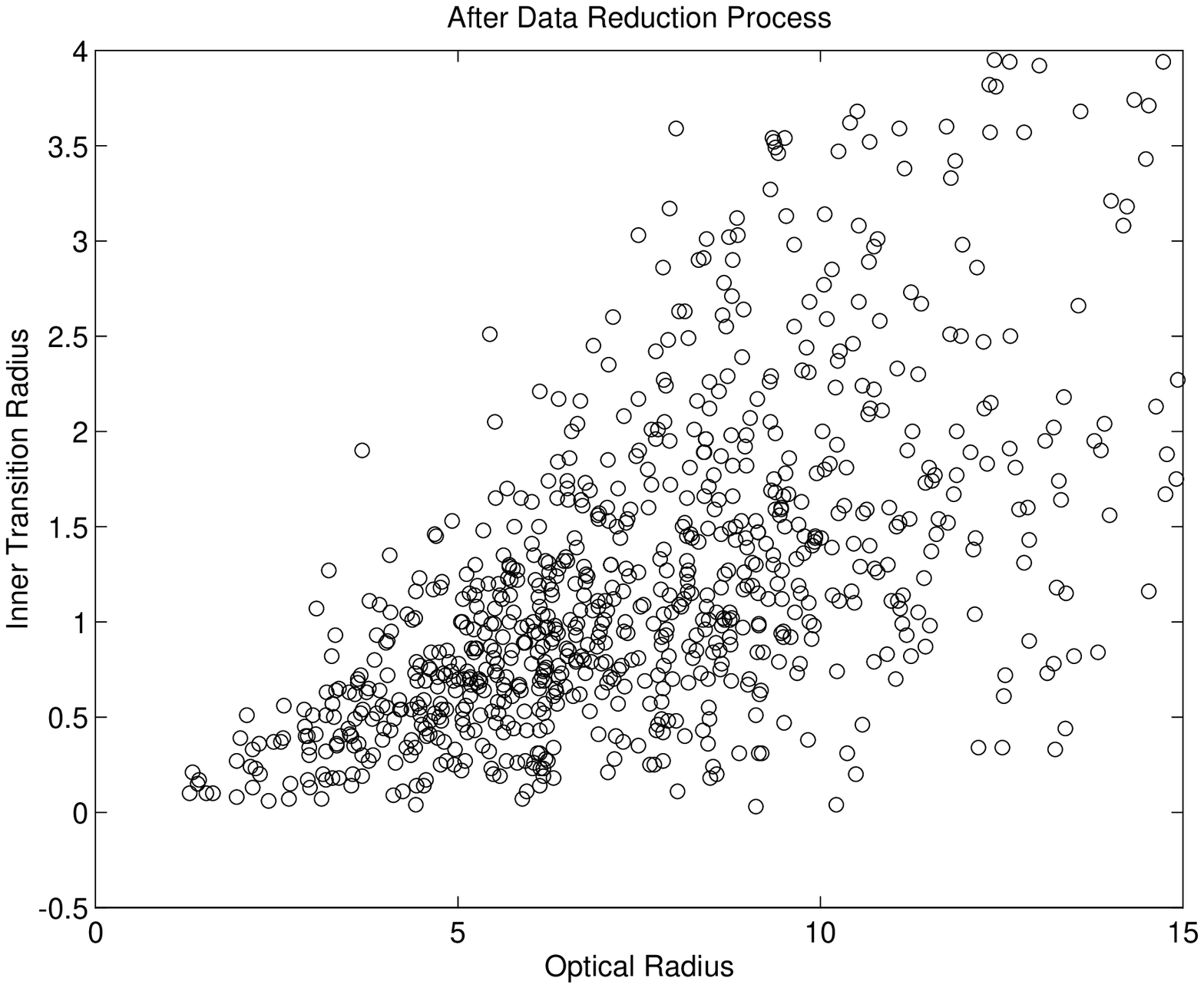}
\caption{After Data Reduction}
\label{fig.1b}
\end{center}
\end{figure*}
It is surprising to note that Table 1 clearly shows that $R_{min}$ in
the {\it unreduced} data is already a very significant (if very noisy)
predictor for $R_{opt}$.
However, after the data reduction process, both the significance and the
reliability of $R_{min}$ as a predictor for $R_{opt}$ increase considerably.

If we consider the post data-reduction values of $R_{min}$ as strong
predictors of bulge-radius (which is reasonable), then the strong
post data-reduction correlation between $R_{min}$ and $R_{opt}$ can be
seen as a correlation between bulge radius and optical disc radius.
The fact that a significant correlation between $R_{min}$ and $R_{opt}$
already existed prior to the data reduction can then be interpreted to indicate
that the pre data-reduction $R_{min}$ values were already significant (if
very noisy) predictors of bulge-radius.

There is a further significant implication of the strong post data-reduction
correlation between $R_{min}$ and $R_{opt}$: the correlation implies a strong
causal connection between the two values which, in turn, implies that any
effect the halo has on $R_{opt}$ is necessarily balanced by a corresponding
effect of the halo on $R_{min}$.
In particular, the correlation is inconsistent with the existence of any strong
transition region between disc-dominated and halo-dominated dynamics in the
optical disc - a conclusion already reached by PS 1986, and others and
clear from Figure \ref{fig.0}.
\section{A Basic Correlation \hfill \break
Imposed By Power-Law \hfill \break
Rotation Curves}
\label{sec.4}
\begin{figure*}
\begin{center}
\includegraphics*[width=3.5in,keepaspectratio]{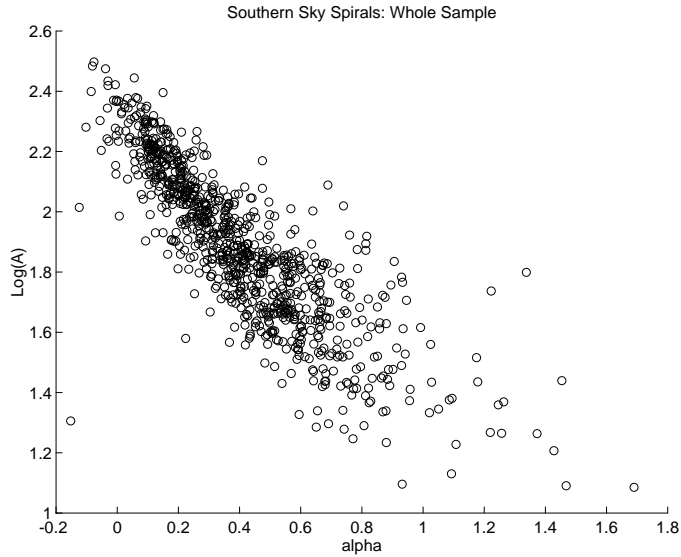}
\caption{$(\alpha,\log A)$ plotted for each of \hfill \break
900 rotation curves}
\label{fig.2}
\end{center}
\end{figure*}

In the following, it is shown that an extremely strong correlation exists
between the power-law parameters over the whole PS sample, and
the correlation is detected in the following way:
\begin{itemize}
\item
The basic assumption is that rotation velocities behave as
$V = A R^\alpha$ so that
$\log V = \log A + \alpha\,\log R$;
\item For each of the 900 rotation curves in the sample, regress
$\log V$ on $\log R$ to obtain estimates of $\log A$ and $\alpha$;
\item Plot the 900 pairs $(\alpha\,,\log A)$ on a single diagram.
\end{itemize}
The results of this exercise are shown in Figure \ref{fig.2},
and shows that there exists an extremely strong negative $(\alpha,\log A)$
correlation.
The only obvious deviations from what could be called an almost perfect
linear correlation are, firstly, a slight hint of curvature along the lower
boundary of the distribution and, secondly, the fan-like behaviour of the
diagram going from a broad spread of points at the bottom right-hand of the
figure to a narrow neck at the top left-hand of the figure.

In the following sections, we show that the structure of Figure \ref{fig.2}
orders the galaxies in the sample according to their absolute magnitude and
surface brightness properties.
\section{The First-Order 70\% Model}
\label{sec.5}
\subsection{The Basic Analysis}
\label{subsec.5.1}
The first-order model simply assumes that the $(\alpha,\log A)$ relationship
underlying Figure \ref{fig.2} is linear, so that
\begin{equation}
\log A = a_0 + b_0 \alpha,
\label{5}
\end{equation}
where $a_0$ and $b_0$ are constants, and linear regression shows that this
model can account for 70\% of the variation (equivalent to a regression
correlation coefficient of 0.84) in Figure \ref{fig.2}.
\subsection{Geometric Implications}
\label{sec.5.1}
It is easily shown that when the parameters, $(m,c)$, of a set of lines
$y=mx+c$ in the $(x,y)$-plane are constrained to satisfy $c=y_0-x_0 m$, for
constants $(x_0,y_0)$, then the lines themselves are constrained to meet at the
fixed point $(x_0,y_0)$ in the plane.
Consequently, in the present context, the first-order linear approximation
(\ref{5}) implies that all rotation curves,
\begin{equation}
\log V = \log A + \alpha \log R,
\label{4a}
\end{equation}
for which $\log A$ and $\alpha$ are related by (\ref{5}), intersect at the
fixed point $(-b_0,a_0)$ in the $(\log R,\log V)$ plane.
If this fixed point is denoted as $(\log R_0,\log V_0)$, then
(\ref{5}) is more transparently written as
\begin{equation}
\log A = \log V_0 - \alpha \log R_0
\label{6}
\end{equation}
so that (\ref{4a}) implies
\begin{equation}
V = V_0 \left( { R \over R_0 } \right)^\alpha.
\label{4b}
\end{equation}
Assuming the model (\ref{6}), a linear regression on the data of Figure
\ref{fig.2} subsequently gives $(\log R_0,\,\log V_0) \approx (0.84,2.22)$.
Whilst the idea of a single intersection point for all rotation curves in the
sample seems rather extreme, it is unambiguously deduced as a consequence
of the first-order modelling of Figure \ref{fig.2} - and is therefore to be
seen as a first-order approximation of the reality.

In the following sections, we give two powerful illustrations of the reality
of the idea that optical rotation curves can be considered, in the first
approximation, to converge in a very small region of the $(\log R,\,\log V)$
plane.
\subsection{The Intersection Frequency \hfill \\ Diagram}
\label{sec.6}
\begin{figure*}
\begin{center}
\includegraphics*[width=3.5in,keepaspectratio]{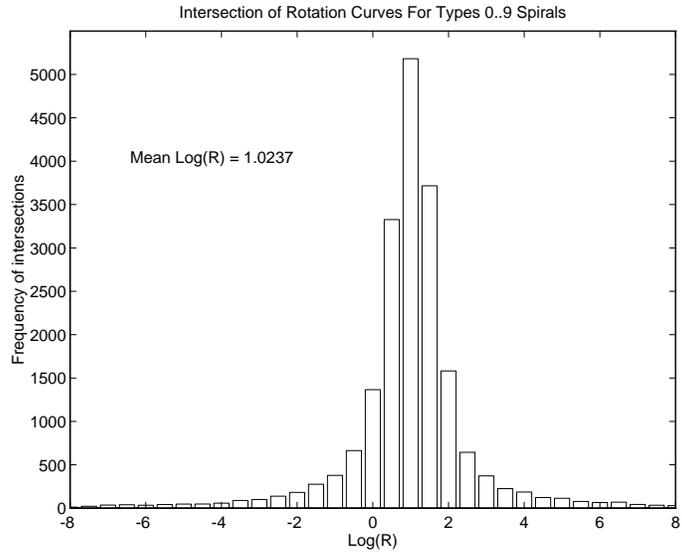}
\caption{Frequency of Intersections on $\log R$ axis}
\label{fig.4}
\end{center}
\end{figure*}

\begin{figure*}
\begin{center}
\includegraphics*[width=3.5in,keepaspectratio]{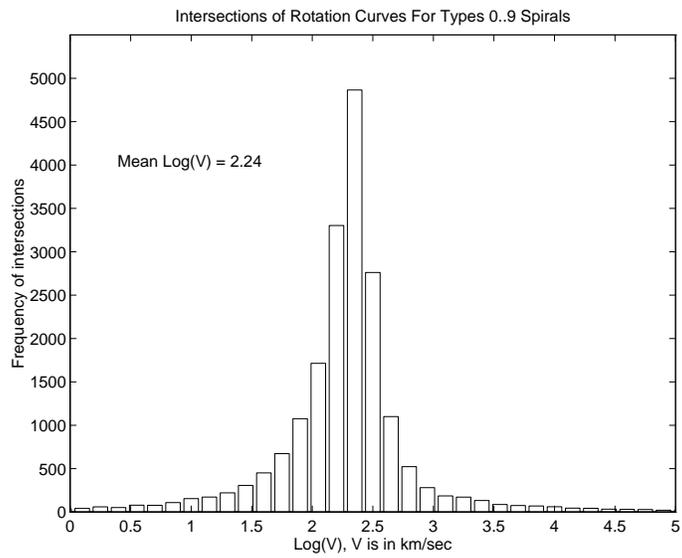}
\caption{Frequency of Intersections on $\log V$ axis}
\label{fig.5}
\end{center}
\end{figure*}

A direct illustration of rotation curve convergence is given as follows: we
have calculated the coordinates of intersection between all possible pairs of
200 randomly sampled rotation curves drawn from the full sample, and have
plotted the frequencies of intersection along the $\log R$ and $\log V$ axes
respectively in Figures \ref{fig.4} \& \ref{fig.5}.
It is clear from these figures that there is a very sharp peak of
intersection points in both diagrams.
Using the mean points in each diagram, together, they are consistent
with a peak intersection point at $(\log R,\log V) \approx (1.02,2.24)$.
\subsection{A Geometric Illustration Of Rotation Curve Convergence}
\label{sec.7}
A second, and dramatic, illustration of rotation curve convergence can be formed
from the realization that, if the rotation curves really do converge on a single
fixed point in the $(\log R,\log V)$ plane, then all of the rotation curves will
{\it transform into each other} under rotations about the fixed point,
$(\log R_0,\log V_0)$, in this plane.

This idea can be tested in the following way: suppose that an arbitrarily
chosen straight line passing through the point $(\log R_0,\log V_0)$ is
defined as a standard {\lq reference line'}.
Since, according to the first-order model interpretation of Figure \ref{fig.2},
all the rotation curves in the sample pass through the point
$(\log R_0,\log V_0)$, then every rotation curve in the sample can be
transformed into the standard reference line by a simple bulk-rotation about
$(\log R_0,\log V_0)$.
Since, according to this idea, the rotation curves are reduced to equivalence
by the rotation, then the process of forming an {\lq average rotation curve'}
from the set of rotated such curves should greatly reduce the internal noise
associated with the individual rotation curves, and we would expect the
resulting average curve to be a very close fit to the standard reference
line, referred to above.

In the following analysis, which is performed for all 900 galaxies in
the sample, the technique described in Appendix \ref{app.4a} is used to
refine the histogram estimate of $(1.02,2.24)$, given in the last section,
to the optimal value for the fixed point as
$(\log R_0,\,\log V_0)=(1.06,2.29)$.
\begin{figure*}
\begin{center}
\includegraphics*[width=3.5in,keepaspectratio]{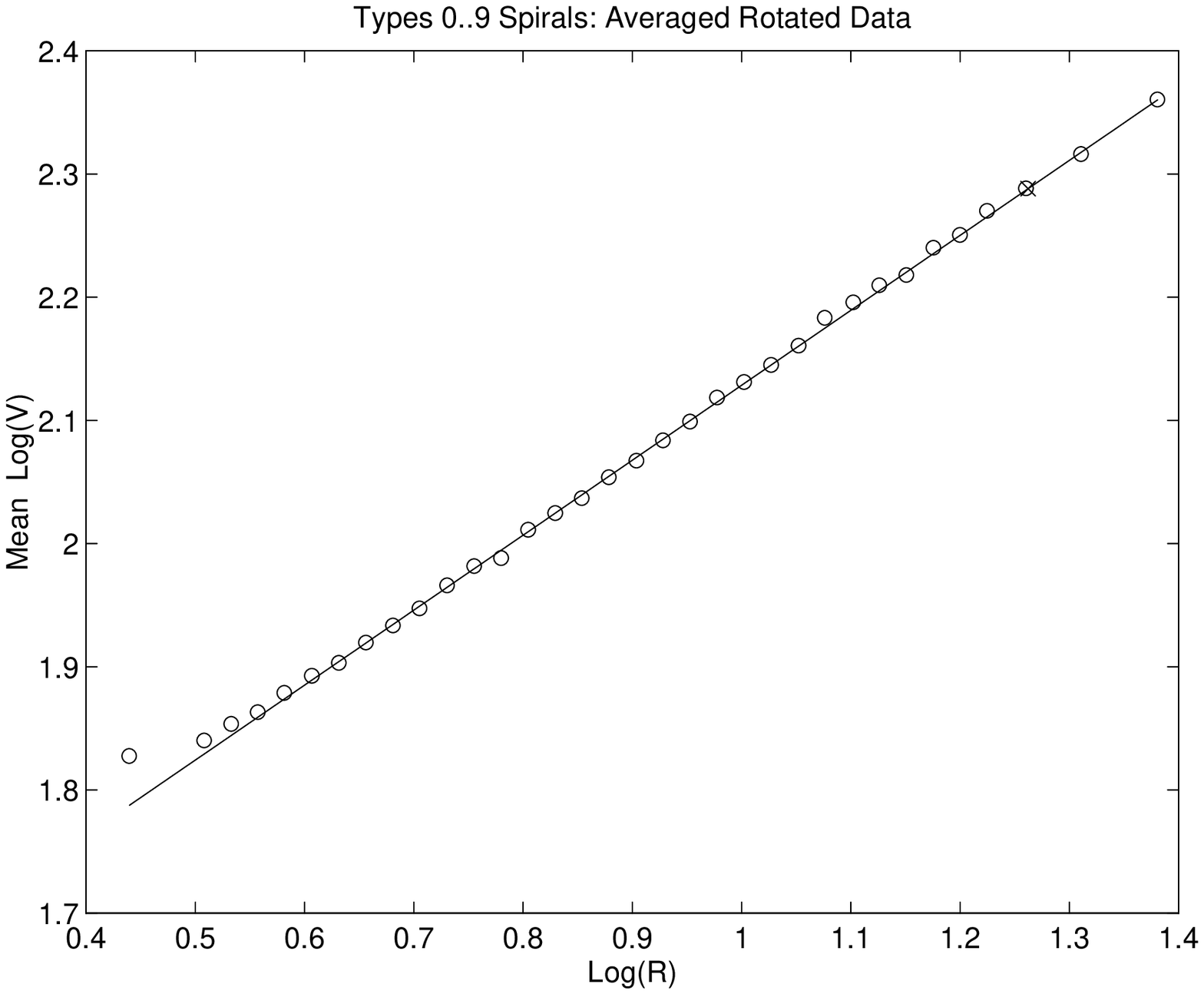}
\caption{Types 0..9 spirals; Averaged rotated data}
\label{fig.7}
\end{center}
\end{figure*}
The solid line in Figure \ref{fig.7} is the standard reference line which
is defined to pass through the fixed point
$(\log R_0,\,\log V_0)=(1.06,2.29)$,
and fixed arbitrarily so that $\log A=1.5$.
The results are not sensitive to the choice of this latter parameter.
The circles in the figure show the results of rotating the 900 rotation curves
of the sample about the fixed point so that they coincide {\it in a least-square
sense} with the standard reference line, and then averaging over all the
rotation curves.
Each circle represents an average of approximately 800 separate measurements.

When Figure \ref{fig.7} is compared with Figure \ref{fig.0}, we see that
virtually all of the scatter present within the latter figure is eliminated
in Figure \ref{fig.7}, thereby providing the strongest possible evidence for
the idea of the equivalence of rotation curves with respect to rotation about
particular fixed points in the $(\log R,\log V)$ plane.
\subsection{A Comparison of Three \hfill \\
$(\log R_0,\, \log V_0)$ Estimates}
As we pointed out in \S\ref{subsec.5.1}, the first-order approximation of the
$(\alpha,\,\log A)$ plot of Figure \ref{fig.2} implies directly that all
$H_\alpha$ rotation curves pass through a single fixed point in the
$(\log R,\,\log V)$ plane, and we have three distinct ways of estimating
the position of this fixed point.
The first way was by a direct linear regression on the data of Figure
\ref{fig.2}, the second way was by estimating the peaks of the intersection
histograms of Figures \ref{fig.4} and \ref{fig.5} respectively, whilst
the third way was by the minimization procedure described in \S\ref{sec.7}
the results of which are shown in Figure \ref{fig.7}.
The three estimates are shown together in Table 4.
\begin{center}
\begin{tabular}{||r|c|c||}
\hline
\hline
\multicolumn{3}{||c||}{\bf Table 4}  \\
\hline
\hline
Method & $\log R_0$ & $\log V_0$  \\
\hline \hline
Regression & 0.84 & 2.22 \\
Mean Values & 1.02 & 2.24 \\
Minimization & 1.06 & 2.29 \\
\hline
\hline
\end{tabular}
\end{center}
There is clearly a high degree of consistency between the three methods which
serves to emphasize the reality of the first-order effect.
\section{The Second-Order 85\% \hfill \\Model}
\subsection{The Basic Analysis}
\label{subsec.6.1}
Whilst, so far, Figure \ref{fig.2} has been analysed on the basis of a
first-order linear model, there is the possibility that systematic internal
structure exists in Figure \ref{fig.2}.
In this section, we show how the existence of this internal structure is
strongly supported on the data, and leads to a second-order model which accounts
for over 85\% of the variation (equivalent to a regression correlation
coefficient of 0.92) in the pivotal diagram, Figure \ref{fig.2}.
\begin{center}
\begin{tabular}{||c|c|c|c||}
\hline
\hline
\multicolumn{4}{||c||}{\bf Table 5}  \\
\hline
\hline
\multicolumn{4}{||c||}{$(\alpha,\,\log A)$ data averaged in} \\
\multicolumn{4}{||c||}{magnitude-limited quartiles} \\
\hline
\hline
Sample  & Mean      & Mean       & Mean      \\
Size    & abs(mag)  &  $\alpha$  &  $\log A$  \\
\hline
\hline
224 & -22.20  & 0.22 & 2.15 \\
227 & -21.16  & 0.34 & 1.96 \\
224 & -20.27  & 0.47 & 1.79 \\
225 & -18.80  & 0.58 & 1.65 \\
\hline
\hline
\end{tabular}
\end{center}

\begin{figure*}
\begin{center}
\includegraphics*[width=3.5in,keepaspectratio]{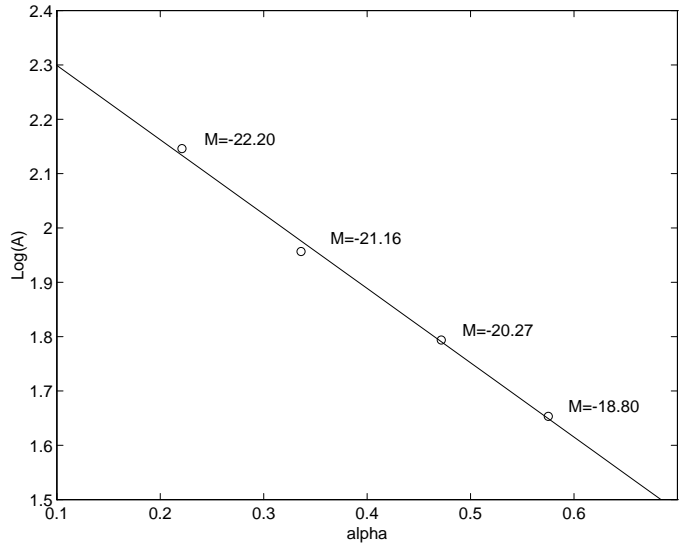}
\caption{$(\alpha,\log A)$ data averaged in magnitude-limited quartiles}
\label{fig.7a}
\end{center}
\end{figure*}

We begin by
taking the 900 pairs of $(\alpha,\log A)$ data which form the basis of
Figure \ref{fig.2} and partition them as near as possible into quartiles
according to absolute magnitude; thus the data is partitioned into the brightest
25\% of the sample, the next brightest 25\% and so on.
For each quartile, we then form the mean values of the absolute magnitude,
$\log A$ and $\alpha$ respectively, and the results of this exercise
are listed in Table 5 and plotted in Figure \ref{fig.7a}.
The solid line in the figure is the linear regression line fitted to the
four data points, and the whole diagram can be considered as a condensed
representation of the data plotted in Figure \ref{fig.2} showing the variation
of absolute magnitude through that data.
It is clear from Table 5 and Figure \ref{fig.7a} that $\alpha$ and $\log A$
are each very strongly correlated with the absolute magnitude, thereby
justifying a more detailed investigation.
\subsection{The Refined Analysis}
\label{subsec.6.2}
\begin{figure*}
\begin{center}
\includegraphics*[width=3.5in,keepaspectratio]{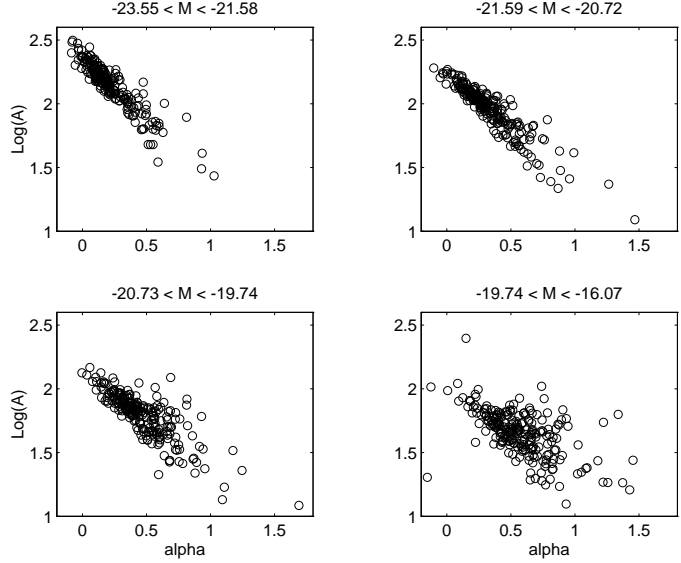}
\caption{$(\alpha,\,\log A)$ plotted for each magnitude-limited quartile}
\label{fig.8}
\end{center}
\end{figure*}

\begin{figure*}
\begin{center}
\includegraphics*[width=3.5in,keepaspectratio]{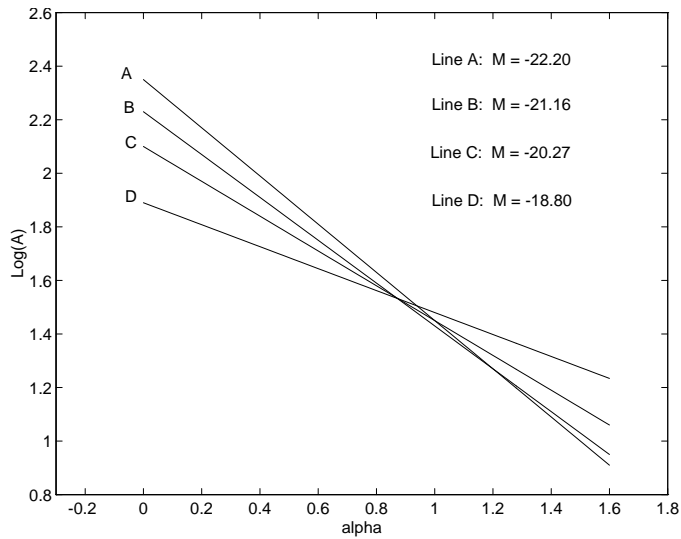}
\caption{Regressions on $(\alpha,\,\log A)$ data in magnitude-limited quartiles}
\label{fig.9}
\end{center}
\end{figure*}

We begin by plotting the $(\alpha,\,\log A)$ data for each of these
magnitude-limited quartiles and displaying these plots in Figure \ref{fig.8}.
Viewed collectively, the four diagrams of this figure display three significant
features:
\begin{itemize}
\item
Firstly, the diagram corresponding to the brightest magnitude-limited
quartile shows that $(\alpha,\,\log A)$ data exhibits almost scatter-free
linear behaviour, whilst the data of the remaining quartiles exhibits
increasing scatter as absolute magnitude increases.
\item
Secondly, apart from this scatter, there appears to be a systematic variation
of the data through the four diagrams when they are viewed in order of
increasing absolute magnitude.
\item
Thirdly, the scatter evident in Figure \ref{fig.2} is seen to be a function of
increasing absolute magnitude
which, at face value, would suggest it is a function of increasing measurement
uncertainties at the dimmer end of the sample.
However, there is a feature which suggests that the situation is not this
straightforward:
specifically, the lower boundary of points in each diagram of Figure \ref{fig.8}
is very sharply
delineated whilst it is the upper boundary which becomes increasingly diffuse
in the higher magnitude quartiles.
If measurement uncertainty is the basic cause of scatter in these diagrams,
then it is not immediately obvious why the lower boundary of points in every
quartile is sharply delineated.
\end{itemize}
The reality of a systematic variation with increasing magnitude in the
diagrams of Figure \ref{fig.8} is confirmed by forming the linear regression
of $\log A$ on $\alpha$ for the data of each of these diagrams so that, for
each diagram, there is a linear model analogous to (\ref{6}).
The results are listed in Table 6 and plotted in Figure \ref{fig.9}.
The figure shows very clearly that the four regression lines meet in a very
small neighbourhood of the point $(0.9,1.5)$ of the $(\alpha,\,\log A)$ plane.
\begin{center}
\begin{tabular}{||c|c|c|c||}
\hline
\hline
\multicolumn{4}{||c||}{\bf Table 6}  \\
\hline
\hline
\multicolumn{4}{||c||}{$\log A = \log V_0 - \alpha\, \log R_0$} \\
\hline
\hline
Sample  & $\log V_0$ & $\log R_0$ & Mean $M$\\
Size    &        &       & of quartile \\
\hline \hline
224 & 2.35 & 0.90 & -22.20 \\
227 & 2.23 & 0.80 & -21.16 \\
224 & 2.10 & 0.65 & -20.27 \\
225 & 1.89 & 0.41 & -18.80 \\
\hline
\hline
\end{tabular}
\end{center}
We have already noted that, when a set of lines $y=mx+c$ in the $(x,y)$-plane
are constrained to meet at a fixed point, $(x_0,y_0)$ say, then the parameters
$(m,c)$ of the lines are constrained to satisfy $c=y_0-x_0 m$.
In the present case, representing any of the lines in Figure \ref{fig.9}
as $\log A = \log V_0 - \alpha\,\log R_0$, this means that the parameters
$(\log R_0,\, \log V_0)$ are
constrained to satisfy $ \log V_0 \approx 1.5 + 0.9\,\log R_0$.
Since, in Figure \ref{fig.9}, the only thing which varies between the lines
is the absolute magnitude, then we can expect $\log V_0$ and $\log R_0$ to
be individual functions of absolute magnitude and this is confirmed by Table 6.
Linear regression of the data of this table gives
\begin{eqnarray}
\log V_0 &=& -0.67 -0.14 \,M, \label{5a} \\
\log R_0 &=& -2.34  - 0.15 \,M \nonumber
\end{eqnarray}
from which we get $ \log V_0 = 1.51 + 0.93 \,\log R_0$, which confirms the
result obtained directly from Figure \ref{fig.9}.
We now note that the constants and coefficients in (\ref{5a}) can be
{\lq fine tuned'} in the following way:
If we substitute equations (\ref{5a}) into (\ref{6}) we obtain the relationship
\begin{eqnarray}
\log A &=& \log V_0  - \alpha\,\log R_0 ~~\rightarrow
\nonumber \\
\log A &=& -0.67 -0.14\,M \nonumber \\
 &+& 2.34\,\alpha  + 0.15\,\alpha\,M,
\label{5b}
\end{eqnarray}
which shows directly that the model (\ref{5a}) is equivalent to a model for
$\log A$ expressed in terms of $M$, $\alpha$ and $\alpha\,M$.
Given this model for $\log A$, its coefficients can be optimised in a
least-square sense by regressing $\log A$ directly on the predictors
$M$, $\alpha$ and $\alpha\,M$ over the whole data set.
This process gives
\begin{eqnarray}
\log A &=& -0.56 -0.13\,M \nonumber \\
&+& 2.27\,\alpha + 0.14\,\alpha\,M,~~
\rightarrow \nonumber \\
\log A &=& \log V_0 - \alpha \, \log R_0,~~{\rm where} \nonumber \\
\log V_0 &=& -0.56 -0.13 \,M, \nonumber \\
\log R_0 &=& -2.27  - 0.14 \,M. \nonumber
\end{eqnarray}
with very strong statistics on the constant and coefficients, and shows that
this regression accounts for over 85\% of the variation (equivalent to a
regression correlation coefficient of 0.92) in the pivotal
diagram, which is Figure \ref{fig.2}.
A comparison with (\ref{5b}) shows only
minor adjustments to the constants and coefficients of the
$(\log R_0,\,\log V_0)$ models.
Consequently, and using (\ref{4b}), the optimised second-order model gives the
universal rotation curve as
\begin{eqnarray}
{ V \over V_0 } &=& \left( { R \over R_0 } \right)^\alpha, \label{7a} \\
\log V_0 &=& -0.56 -0.13 \,M, \nonumber \\
\log R_0 &=& -2.27  - 0.14 \,M. \nonumber
\end{eqnarray}
\section{The Third-Order 90\% Model}
It has been shown that over 85\% of the variation in the pivotal diagram,
Figure \ref{fig.2}, is accounted for by the second-order model (\ref{7a})
which depends only on absolute magnitudes.
However, the PS data-base for the 900 rotation curves also provides estimates
to the optical radius, $R_{opt}$, which, with the absolute magnitudes, allows
us to estimate the  surface brightness of each galaxy in the sample;
since this represents independent information, it seems sensible to consider
the effect of its inclusion in the model and, for present purposes, we define
it to be in solar luminosities per square parsec so that
\begin{displaymath}
S \approx \left( { 2.5^{5-M} \over \pi R_{opt}^2 } \right).
\end{displaymath}

The second-order model, (\ref{7a}), was arrived at by initially noting that
Table 6 implied the relations (\ref{5a}) which, in conjunction with the basic
relationship $\log A = \log V_0 - \alpha\,\log R_0$, allowed the deduction of the
structure
\begin{displaymath}
\log A = a_0 + a_1 \,\alpha + a_2 \, M + a_3\, \alpha \, M
\end{displaymath}
for the $\log A$ model.
If (\ref{7a}) is to be modified by the introduction of $S$, so that
\begin{eqnarray}
\log V_0 = b_0 + b_1\,M + b_2\, S \nonumber \\
\log R_0 = c_0 + c_1\,M + c_2\, S, \nonumber
\end{eqnarray}
then the $\log A$ model will have the structure
\begin{eqnarray}
\log A &=& b_0 + b_1 \,M + b_2 \, S \nonumber \\
&-& c_0\,\alpha - c_1\, \alpha\,M - c_2\,\alpha \, S. \label{6a}
\end{eqnarray}
The results of regressing $\log A$ on the predictors $M$, $S$, $\alpha$,
$\alpha\,M$ and $\alpha\, S$ are given in Table 7.
\begin{center}
\begin{tabular}{||l|r|r|r|c||}
\hline
\hline
\multicolumn{5}{||c||}{\bf Table 7}  \\
\hline
\hline
\multicolumn{5}{||c||}{$
\log A = b_0 + b_1 \,M + b_2 \, S - c_0\,\alpha - c_1\, \alpha\,M -
c_2\,\alpha \, S$}   \\
\hline
\hline
Pred  & Coeff          & Std Dev        & t-ratio & p \\
           & $\times\,10^4$ & $\times\,10^4$ &         &   \\
\hline \hline
Const.       & -5840 &  890                   & -7 & 0.00 \\
$M$          & -1330 &  44                    & -30 & 0.00 \\
$S$          & -2.434 & 0.77                 & -3 & 0.00  \\
$\alpha$     & 32910 & 1600                 & 21 & 0.00 \\
$\alpha\,M$  & 2080 &  84                   & 25 & 0.00 \\
$\alpha\, S$ & 29.23 & 2                    & 14& 0.00  \\
\hline
\multicolumn{5}{||c||}{ $R^2 ~=~90.3\%$} \\
\hline
\hline
\end{tabular}
\end{center}
This model now accounts for marginally over 90\% of the variation
(equivalent to a regression correlation coefficient of 0.95) in the pivotal
diagram, Figure \ref{fig.2}, and we see from the t-ratios that the surface
brightness is most certainly a significant component of the model.
Interpreting the table according to the model
\begin{eqnarray}
\log A = \log V_0 - \alpha\, \log R_0, \nonumber \\
\log V_0 = b_0 + b_1\,M + b_2\, S, \nonumber \\
\log R_0 = c_0 + c_1\,M + c_2\, S, \nonumber
\end{eqnarray}
we arrive, finally, at the third-order model
\begin{eqnarray}
{ V \over V_0 } &=& \left( { R \over R_0 } \right)^\alpha, \label{7b} \\
\log V_0 &=& -0.584 - 0.133 \,M - 0.000243\, S, \nonumber \\
\log R_0 &=& -3.291 - 0.208 \,M - 0.00292\, S. \nonumber
\end{eqnarray}
Referring to Table 7 again, we see that $S$ is a very strong predictor for
$\log R_0$, whilst its effect on $\log V_0$, whilst significant, is much less so.
This conclusion is supported by numerical experimentation which shows that it
can be omitted from $\log V_0$ with hardly any effect on the model.
However, Table 7 shows that it probably is a real predictor for $\log V_0$,
and so it is retained in the model.
\subsection{A Secondary Test Of The Model}
\begin{figure*}
\begin{center}
\includegraphics*[width=3.5in,keepaspectratio]{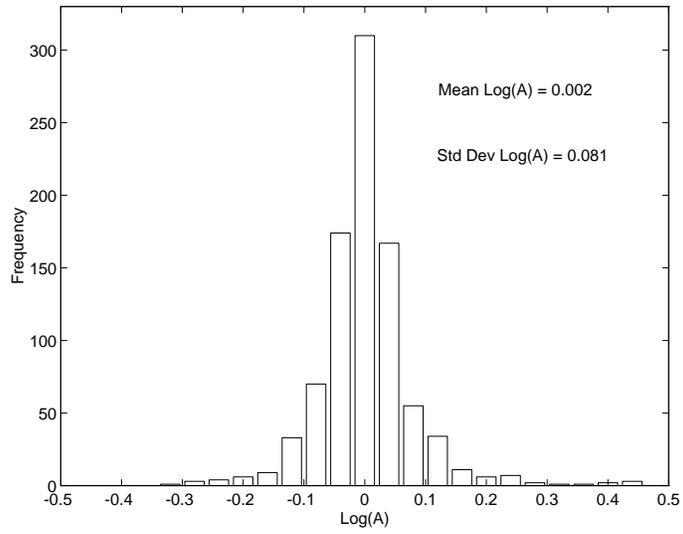}
\caption{Distribution of 900 regression constants}
\label{fig.10}
\end{center}
\end{figure*}

A secondary test of the validity of the third-order model can be given
as follows: For each galaxy we compute $(R_0,V_0)$
according to (\ref{7b}), form the scaled profile $(R/R_0,\,V/V_0)$
and regress $\log (V/V_0)$ on $\log (R/R_0)$.
If (\ref{7b}) is good enough, then the regression constant should be
{\it statistically} zero.
The distribution of the 900 regression constants is shown in Figure \ref{fig.10}, for which a one-parameter t-test gives the 95\% confidence
interval for the mean of the distribution as $(-0.003,+0.007)$ whilst the
99.999\% confidence interval is given as $(-0.010,+0.014)$.
Thus, for all practical purposes, the regression constant can be considered
as statistically zero, thereby confirming the quality of the model.
\subsection{The Brightest Quartile}
\begin{figure*}
\begin{center}
\includegraphics*[width=3.5in,keepaspectratio]{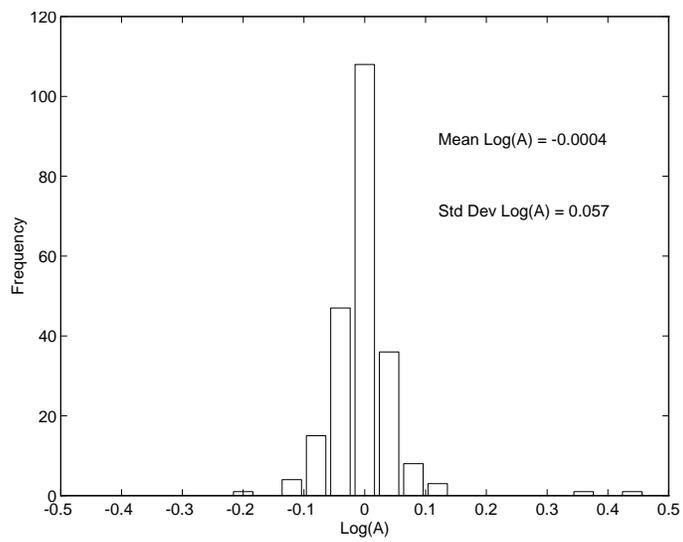}
\caption{Brightest quartile regression constants}
\label{fig.11}
\end{center}
\end{figure*}

Finally, as a means of demonstrating that the (already small) scatter about
the zero point evident in Figure \ref{fig.10} is a function of uncertainty in
the data, rather than due to a poor fit of the basic power-law model, we apply
the third-order model directly to the 224 rotation curves of the brightest
quartile, for which the $(\alpha,\,\log A)$ data is plotted in the first diagram
of Figure \ref{fig.8}.
For this data, the distribution of the 224 regression constants arising from the
linear regression of $\log (V/V_0)$ data on $\log (R/R_0)$ data is given in
Figure \ref{fig.11}.
A direct visual comparison with Figure \ref{fig.10} shows immediately that the
scatter of points about the zero point is very much reduced in the brightest
quartile data, thereby supporting the hypothesis that the scatter which exists
in Figures \ref{fig.10} and \ref{fig.11} is due purely to uncertainty in the
data, and not at all due to any lack of fit of the general power-law model.
\subsection{The Surface-Brightness \hfill \break Cut-Off}
It has already been noted that the lower boundary of points in each of
the magnitude-limited quartile plots of $(\alpha,\, \log A)$ shown in
Figure \ref{fig.8} is very sharply delineated, whilst the corresponding
upper boundaries become increasingly diffuse as magnitude increases.
We shall show that this phenomenon arises because the lower boundary of
points in each of this diagrams effectively defines a {\it zero}
surface-brightness boundary and therefore represents a real physical
cut-off boundary.
By contrast, the upper boundaries are closely associated with maximum surface
brightnesses in each of the quartiles, and so the magnitude-dependent scatter
observable through the four magnitude-limited quartiles almost certainly
arises from observational uncertainties.
\begin{figure*}
\begin{center}
\includegraphics*[width=3.5in,keepaspectratio]{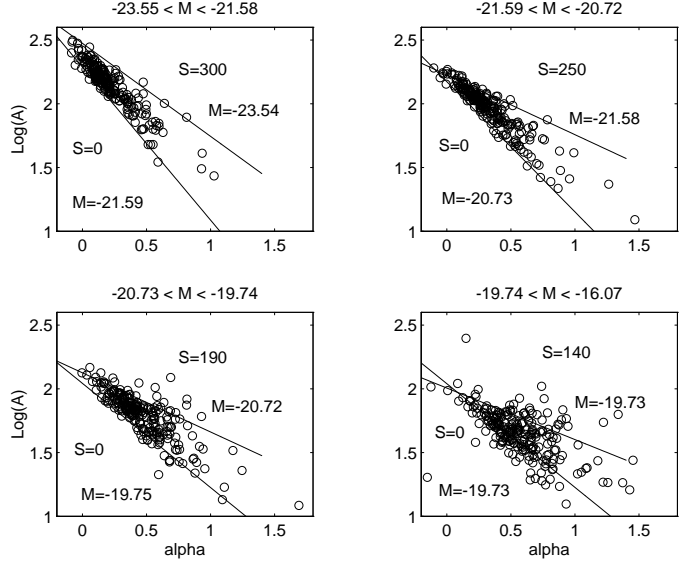}
\caption{Surface-brightness boundaries shown for each magnitude-limited
quartile}
\label{fig.13}
\end{center}
\end{figure*}
This is shown in the following manner: for each of the quartiles, which is
magnitude-limited in its own specific range $(M_{min} < M < M_{max})$,
we determine the surface-brightness limits, $(0 < S < S_{max})$, which
contain 95\% of the sample (to eliminate the effects of long tails).
Then, using the third-order model, $\log A = F(\alpha,\,M,\, S)$, we find that
the distribution of points in each of the quartiles is closely bounded by
the lines
\begin{eqnarray}
\log A &=& F(\alpha,\, M_{max},0), \nonumber \\
\log A &=& F(\alpha,\, M_{min},S_{max}), \nonumber
\end{eqnarray}
in the $(\alpha,\,\log A)$ plane for each of the magnitude-limited quartiles.
The results are shown in Figure \ref{fig.13}, and the plots in the three
brightest quartiles show very clearly that the lower boundaries of points
in these quartiles coincide very closely with the $(S=0,\,M= M_{max})$
boundary in the plane, whilst the diffuse upper boundaries of points coincide
very closely with the $(S=S_{max},\,M=M_{min})$
boundary in the plane.

The only partial exception to this rule is the plot in the dimmest
magnitude-limited quartile.
Here, the upper (very diffuse) boundary of points still coincides
with the $(S=S_{max},\,M=M_{min})$ boundary; however, it was found that the
$(S=0,\,M=M_{max})$ boundary is not even close to the lower boundary of
points in the plot but, instead, we find that the lower-boundary of points
follows closely the $(S=0,\,M=M_{min})$ boundary.
The important point is that the lower boundary of points is still a
$S=0$ boundary.
The obvious explanation for this discrepancy is that objects of very low
surface brightness and very high magnitude are observationally excluded by
selection effects, and therefore do not exist in the dimmest quartile
sample whilst, simultaneously, the same selection effects will also ensure that the
low surface-brightness objects in the dimmest quartile will tend to be the
low magnitude objects.

To summarize, the sharply delineated lower boundaries in the magnitude-limited
quartiles of Figures \ref{fig.8} or \ref{fig.13} define a physical $S=0$ cut-off
boundary, whilst the increasingly diffuse upper boundaries probably arise as a
function of the increasing observational uncertainties associated with
increasing magnitudes.
\section{A Comparative Test Of \hfill \break
The First, Second And\hfill \break
Third-Order Models}
\begin{figure*}
\begin{center}
\includegraphics*[width=3.5in,keepaspectratio]{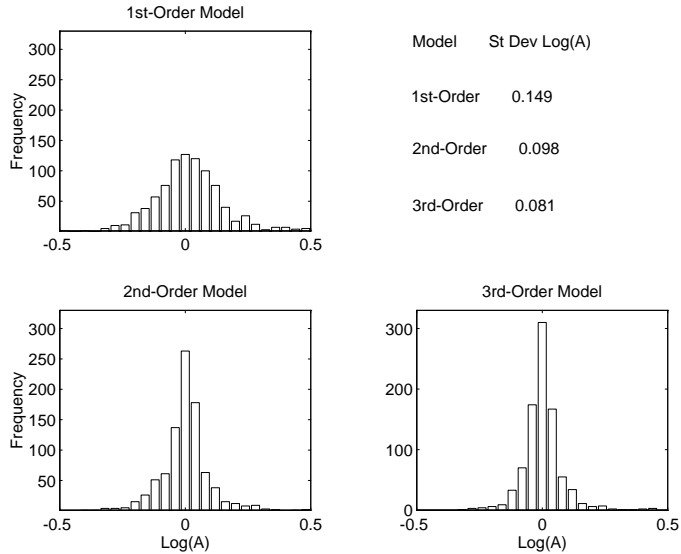}
\caption{Comparative test of the three models}
\label{fig.12}
\end{center}
\end{figure*}

The three models share the basic structure
\begin{equation}
{V \over V_0} = \left( { R \over R_0 } \right)^\alpha
\label{7c}
\end{equation}
and are distinguished only by the models assumed for $(R_0,\,V_0)$.
These three models are defined, respectively, according to:
\hfill
\leftline{The First-Order Model}
\begin{eqnarray}
\log R_0 &=& 1.02, \label{7d} \\
\log V_0 &=& 2.29;  \nonumber
\end{eqnarray}
\leftline{The Second-Order Model}
\begin{eqnarray}
\log R_0 &=& -2.27  - 0.14 \,M, \label{7e} \\
\log V_0 &=& -0.56 -0.13 \,M; \nonumber
\end{eqnarray}
\leftline{The Third-Order Model}
\begin{eqnarray}
\log R_0 &=& -3.291 - 0.208 \,M - 0.00292\, S, \label{7f} \\
\log V_0 &=& -0.584 - 0.133 \,M - 0.000243\, S. \nonumber
\end{eqnarray}
As a final comparative test of these three models we note that, assuming
each to be {\lq correct'}, then a linear regression of $\log (V/V_0)$ data
on $\log (R/R_0)$ data should give rise to a statistically zero constant
terms, indicating that the regression lines (tend to) pass through the origin.
The distribution of these constant terms for each of the 900 rotation curves
is given, for each of the three models, in Figure \ref{fig.12}.
To make the visual comparison easy, the three diagrams in the figure are each
drawn to the same scale, and it is clear that the process of successively
refining the model has the effect of tightening the distribution of $\log A$
about the zero-point.
Specifically, the first-order model is comparatively very poor, whilst
the third-order model is visually significantly better than the second-order
model.
\begin{center}
\begin{tabular}{||r|r|r||}
\hline
\hline
\multicolumn{3}{||c||}{\bf Table 8}  \\
\hline
\hline
\multicolumn{3}{||c||}{One-parameter t-test for confidence} \\
\multicolumn{3}{||c||}{limits of Mean($\log A$)} \\
\hline
\hline
Model  & 95\% Interval & 99.999\% Interval\\
\hline \hline
1st-Order & 0.009,0.029& -0.003, 0.041 \\
2nd-Order & -0.006,0.007 & -0.014,0.015 \\
3rd-Order & -0.003,0.007 & -0.010,0.014 \\
\hline
\hline
\end{tabular}
\end{center}
The results of a one-parameter t-test for limits on the position of
the mean value of $\log A$ for each of the models are given in Table 8.
Apart from showing that the $\log A$ distribution for the 1st-order model
is slightly asymmetric with respect to the zero point, the table shows
that the 2nd and 3rd-order models place very tight 95\% and 99.999\%
confidence intervals on the mean of $\log A$ about the zero point.
Thus, for all practical purposes, the regression constant computed over the
900 rotation curves is statistically zero in both the second and third-order
models, so that these models receive the strongest possible support from the
data.
\section{Equivalence Classes Of Rotation Curves}
\label{sec.10}
As already discussed in detail in \S\ref{sec.5}, the first-order model, equation
(\ref{7c}) with (\ref{7d}), effectively defines the set of all rotation
curves in the sample as a single class within which all the rotation curves
pass through the single point $(\log R_0,\, \log V_0)$ in the $(\log R,\,\log V)$
plane.

By contrast, the second-order model, equation (\ref{7c}) with (\ref{7e}),
replaces the single point, $(\log R_0,\, \log V_0)$, of the first-order model
by a {\it class} of such points parametrized by $M$.
All rotation curves in the class specified by $M$ then pass through the single
point, $(\log R_0,\, \log V_0)$, defined by (\ref{7e}).

The third-order model, equation (\ref{7c}) with (\ref{7f}), is a refinement of
the second-order model in which the class to which any given rotation curve
belongs is defined by specifying $M$ {\it and} $S$, the surface brightness.
All rotation curves in the class specified by $(M,\,S)$ then pass through the
single point, $(\log R_0,\, \log V_0)$, defined by (\ref{7f}).

If we make the (reasonable) assumption that all of the systematic variation
in the pivotal diagram, Figure \ref{fig.2}, has been accounted for by the
90\% third-order model then we can realistically assume that, although the
functional form of the model is probably an approximation to some ideal,
there are no further physical parameters involved in the specification of
$(\log R_0,\,\log V_0)$.
It will then follow that the rotation curves of all galaxies of having the
same absolute magnitude and surfaces brightness will pass through a
single point, $(\log R_0,\,\log V_0)$, in the $(\log R,\,\log V)$ plane.
Consequently, all such rotation curves will be equivalent to within a
rotation through the point, and can therefore be considered to define
an {\it equivalence class} of rotation curves.
\section{The Universal Rotation \hfill \break Curve}
Using the third-order model, it is shown that the universal rotation curve for
spiral galaxies can reasonably be assumed to have the general structure
\begin{eqnarray}
{V \over V_0} &=& \left( { R \over R_0 } \right)^\alpha  \label{7g} \\
\log R_0 &=& -3.291 - 0.208 \,M - 0.00292\, S, \nonumber \\
\log V_0 &=& -0.584 - 0.133 \,M - 0.000243\, S, \nonumber \\
\alpha &=& g (d_m,\,b_m,\,h_m) \nonumber
\end{eqnarray}
where $g (d_m,\,b_m,\,h_m)$ is some undetermined function for which the most
significant parameters are probably the disc-mass,
bulge-mass and halo-mass respectively of the spiral galaxy concerned.
\subsection{The General Situation}
Reference to (\ref{7c}) with (\ref{7d}), (\ref{7e}) and (\ref{7f})
shows that, once a model for $\alpha$ in terms of the physical properties
of spiral galaxies is obtained, then (\ref{7c}) with any one of (\ref{7d}),
(\ref{7e}) or (\ref{7f}) can be considered to represent a universal rotation
curve.
In the following, we consider the problem within the context of the
third-order model.

As we have already noted, the issue depends entirely upon the function
$\alpha$, and what galaxy parameters it is a function of.
\begin{center}
\begin{tabular}{||l|c|c|r|c||}
\hline
\hline
\multicolumn{5}{||c||}{\bf Table 9}  \\
\hline
\hline
\multicolumn{5}{||c||}{$\alpha = b_0 + b_1 \,M $}   \\
\hline
\hline
Predictor  & Coeff & Std Dev & t-ratio & p \\
\hline \hline
Const.    & 2.560 & 0.112  & 23 & 0.00 \\
$M$         & 0.105 &0.005  & 19 & 0.00 \\
\hline
\multicolumn{5}{||c||}{$R^2$ ~=~ 29.2\%} \\
\hline
\hline
\end{tabular}
\end{center}
A least-square linear modelling of $\alpha$ in terms of the available
parameters, $M$ and $S$, gave the best model as $\alpha=-0.583+0.105\,M$,
with no evidence of any significant dependency on $S$.
The details of the best model are given in Table 9 which shows the dependency of
$\alpha$ on $M$ to be extremely significant, but which also shows that the model
accounts for only 29\% of the total variation in the $(\alpha,\,M)$ plot
(equivalent to a regression correlation of approximately 0.55).
To test the possibility that the scatter in this latter plot was, perhaps,
due to noisy data, we tried modelling on the brightest 25\% and 50\% of the
data, but found no evidence at all to support this possibility.
The only reasonable conclusion was that the unaccounted-for 71\% variation
in the $(\alpha,\,M)$ plot arises through the effects of unaccounted-for
additional physical parameters.

The major unincluded determinants of disc dynamics are the relative amounts of
disc-mass, halo-mass and bulge-mass.
Consequently, defining $d_m \equiv$ disc-mass,
$b_m \equiv$ bulge-mass and $h_m \equiv$ halo-mass, and noting that
$M$ is effectively a measure of the total visible mass, then we can
reasonably assert that $\alpha \approx g (d_m,\,b_m,\,h_m)$.
\subsection{A Simple Approximation}
Finally, it is interesting to note how the foregoing analysis allows the
construction of a very simply universal rotation curve, based on absolute
magnitude alone.
This is given by (\ref{7c}) and (\ref{7e}) together with the model for
$\alpha$ described in Table 9 so that, specifically:
\begin{eqnarray}
{V \over V_0} &=& \left( { R \over R_0 } \right)^\alpha  \nonumber \\
\log R_0 &=& -2.27  - 0.14 \,M, \label{7h} \\
\log V_0 &=& -0.56 -0.13 \,M, \nonumber \\
\alpha &=& 2.560+0.105\,M,  \nonumber
\end{eqnarray}
which can be considered as a formal refinement of the information contained
in Figure \ref{fig.7a}.
This latter approximation confirms the conclusions of PS 1996
which are that the universal rotation curve is substantially determined by
absolute magnitude alone, especially for the brightest galaxies.
\section{Theory-Independent Dark-Matter Models}
PSS, used essentially the same data set as that
analysed here to show that whilst, at high luminosities, there is only a slight
discrepancy between observed rotation curves and those predicted on the basis
of the observed luminous matter distributions, the discrepancy is far more
serious at low luminosities.
This conclusion is already strongly supported by the simple model given
at (\ref{7h}) which shows how the power-law exponent, $\alpha$, increases
strongly with absolute magnitude.
Since, (relatively) large values of $\alpha$ correspond directly to the
most steeply rising rotation curves, and since it is precisely these kinds
of rotation curves that require the largest amounts of dark matter for their
explanation, according to the virial theorem, then the findings
of the present analysis are in direct accord with those of PSS.

In this way, PSS have already used the data to show that the dark-matter
component of spiral galaxies does not seem to have the form of an unconstrained
and independent structure added onto the visible structure of spiral galaxies,
but appears to make its presence felt in some kind of systematic way which is
strongly inversely correlated with the visible component of the galaxy
structure.
This effect, together with the detailed considerations of the foregoing
sections, allows the construction of dark-matter models which are
{\it independent} of any particular dynamical theory and therefore, as a
side effect, provide strong tests of such theories.

Specifically, referring to the final form of the universal rotation curve,
given at (\ref{7g}), we see how dark-matter can only make its presence felt
through the exponent, $\alpha$.
Given that reasonable light-based estimates can be made of disc-mass ($d_m$) and
bulge-mass ($b_m$) for spiral galaxies, then it is (in principle) a simple
process to form empirical models of $\alpha$ directly in terms $d_m$ and $b_m$.
Once such models are formed for a reasonably sized data-base of spiral galaxies,
then it becomes a straightforward process to investigate the structure of any
systematic deviations of predictions based on the model
$\alpha \approx g(d_m,\,b_m)$ from the $\alpha$-values calculated directly
from the rotation curves.
The presence of any such systematic deviations can then reasonably be considered
due to the presence of a dark-matter halo, so that, subsequently, estimates
for the amount of dark-matter present for any given galaxy can be given in
terms of the directly measurable properties of the galaxy concerned.
\section{Conclusions}
By demonstrating how completely the power-law model for rotation curves resolves
rotation curve data, magnitude data and surface brightness data over a large
data-base, this paper:
\begin{itemize}
\item substantially refines the well-known correlations that exist between the
rotational kinematics of spiral galaxies and various of their non-kinematic
properties, such as absolute magnitude;
\item shows how the power-law rotation curve is, almost certainly, exact for
idealized discs (that is, for discs without the irregularities inevitably
present in real optical discs).
\end{itemize}
Additionally, at its lowest approximation, the analysis confirms the conclusion
of PSS (1996) that the rotation curve of any given galaxy is largely determined by the
absolute luminosity of the galaxy concerned, and supports their conclusion
that absolute luminosity is inversely correlated with dark mass.
Such an inverse correlation obviously provides a means of dark-matter
modelling, but the detailed nature of the present analysis allows a further
step in providing a means for the generation of statistical dark-matter models
which are independent of any particular dynamical theory.
Apart from the objective desirability of such models, they have an obvious
potential in distinguishing between competing dynamical theories.

Furthermore, in showing that optical discs contain {\it no} signature
of the transition from disc-dominated dynamics to halo-dominated dynamics,
and in showing the existence of a strong correlation between $R_{min}$
and $R_{opt}$, the work provides very strong support for the conclusion
of PS (1986) that the disc-mass distribution and the halo-mass distribution
are very strongly correlated, and suggests that the bulge-mass distribution
should also be factored into this correlation.
\appendix
\section{The Constancy of ${\bar V}$}
\label{app.1b}
PS (1986) showed that the parameter
\begin{displaymath}
{\bar V} \equiv \left( \Omega - {K \over 2} \right) R,
\end{displaymath}
where $\Omega$ is angular velocity, and $K$, the epicyclic frequency, is
defined by
\begin{displaymath}
K^2 ={1 \over R^3} {d \over dR} \left( V^2 R^3 \right),
\end{displaymath}
is almost constant over the larger part of any given optical disc for a
sample of 42 spiral galaxies, and they deduce from this constancy that
rotation curves contain no signature indicating transition from disc-dominated
dynamics to halo-dominated dynamics.
In the present case, the same conclusion can be drawn from the fact that
the kinematics over the whole optical disc are amenable to description by
a {\it single} power-law prescription.
Such a power-law also provides an extra insight into the constancy of
the ${\hat V}$ parameter:

Using the general power-law prescription $V = AR^\alpha$ in the above we
find, after some algebra,
\begin{displaymath}
{ d {\hat V} \over d R} = A\, R^{\alpha-0.5}
\left( {\alpha \over \sqrt{R}} - {1 \over 2}\left(\alpha +{1 \over 2}\right)
\sqrt{2\,\alpha+3} \right).
\end{displaymath}
From this, we see that, whenever whenever a rotation curve is such that
$\alpha < -0.5$, then the corresponding $d {\hat V}/ dR$ rapidly becomes
small for increasing $R$ - and this corresponds to a nearly constant
${\hat V}$.
Reference to Figure \ref{fig.7a} shows that this conditions holds for all
galaxies brighter than $M \approx -20$ and, of the PS (1986) sample of 42
galaxies, 37 are brighter than $M \approx -20$ so that $\alpha < -0.5$
generally, indicating very flat ${\hat V}$ functions, as actually
shown by PS.
\section{Data Reduction}
\label{app.4}
A basic assumption is that optical rotation curves can be described in terms of
a simple power law, and the validity of this assumption as, at the very least,
a good approximation to the reality over the approximate range $(1,20)kpc$
has been demonstrated in \S\ref{sec.4}.

However, spiral galaxies can be considered to consist of a spherical
bulge surrounded by a flat disk, and we can be reasonably certain that
the bulge and the disk represent distinct dynamical regimes.
In practice therefore, the structure of the inner parts of rotation curves will
be determined one dynamical regime, whilst the structure of the outer parts
of the rotation curves will be determined by another.
Since it is not reasonable to suppose that a single simple power law can bridge
the two dynamical regimes, there is a need to find a way of eliminating the
effect of the bulge dynamics from the rotation curve data.
In order to ensure that no subjective bias is imposed on the data during such
a data reduction process, the process adopted must be automatically applicable
according to strictly predefined rules.
The adopted procedure is described below.

The original analysis was done using the commercial statistics package,
{\it Minitab}; when regressions are performed in {\it Minitab}, it
automatically flags observations which have an unusual predictor {\it or}
if they have an unusual response.
In both cases {\lq unusual'} is defined by internal parameters of the
software.
It was decided, in advance of any data processing, that when a {\it Minitab}
regression flagged the {\it innermost} observation on any given rotation
curve (that is, the one most likely affected by the central bulge) as
{\lq unusual'} then that observation would be deleted from the analysis, and
the regression repeated.
If, on the repeated regression, the new innermost observation was flagged as
unusual, then it too would be deleted from the analysis.
This process was repeated until the innermost observation remained unflagged
after the regression process.

There are 900 rotation curves with a total of 19183 observations recorded
with a non-zero radial coordinate (those with a zero radial coordinate
were necessarily deleted because data was transformed into logarithmic form).
Of these 19183 observations, the {\lq innermost deletion'} strategy led to
the exclusion of 2264 observations, which is $11.8 \%$ of the sample.
This was considered to be an acceptable attrition rate.
\section{Determination of Fixed \hfill \break Points,
$(\log R_0,\,\log V_0)$}
\label{app.4a}
In \S4 and \S5 it was determined that the form of Figure \ref{fig.2} implied
the first-order hypothesis that all rotation curves in the sample passed
through a fixed point, $(\log R_0,\log V_0)$, in the $(\log R,\,\log V)$
plane.
The problem is to determine the position of the fixed point
$(\log R_0,\,\log V_0)$.
In the following, we describe how this can be done using a minimization
procedure.

The hypothesis implies that all 900 rotation curves in the sample should be
equivalent to within a rotation about $(\log R_0,\,\log V_0)$; in this case it
should be possible to superimpose all such rotations curves upon each other by
means of a simple appropriate rotation.
However, the very noisy nature of rotation curve data means that, at best, such
a process of superimposition can only be defined in some averaged statistical
way.

In practice, what is done can be characterized as follows:
\begin{itemize}
\item Define an initial guess for the fixed point $(\log R_0,\,\log V_0)$; this
is done by the simple expedient of identifying the peaks of the rotation-curve
intersection frequency distributions show in Figures \ref{fig.4} \& \ref{fig.5},
and they are found to be $\log R \approx 1.02$ and $\log V \approx 2.24$.
\item Define a standard reference curve passing through the estimated fixed
point; typically, this is done by making the arbitrary choice $\log A = 1.5$
and then using (\ref{6}) to determine $\alpha$.
\item Rotate the data corresponding to each individual rotation curve about
this estimated $(\log R_0,\,\log V_0)$ until it coincides {\it in a least-square
sense} with the standard reference curve.
\item Keep a cumulative sum of all the least-square residuals arising from
the foregoing rotation operations performed on the individual rotation curves.
\item Minimize this cumulative sum of least-square residuals with respect
to variations in the estimated position of the fixed point,
$(\log R_0,\,\log V_0)$.
\end{itemize}
Note that the minimization processes referred to above are minimizing functions
defined on noisy data.
It is routinely recommended that such problems are solved using the
Simplex method, which is extremely robust - although slow.
The original implementation of the method is given by Nelder \& Mead
(1965).

\end{document}